\documentclass[preprint,noshowpacs,noshowkeys,endfloats]{revtex4}

\usepackage{dcolumn}
\usepackage{amsmath}
\usepackage{bm}
\usepackage{graphicx}
\usepackage{graphics}
\usepackage{longtable}
\usepackage{color}
\usepackage{epsfig}
\usepackage[normalem]{ulem}

\begin{document}

\title{Magnetization relaxation  and   search for the magnetic gap in bulk-insulating V-doped (Bi, Sb)$_2$Te$_3$ }

\author { E. Golias$^1$,  E. Weschke$^{1}$, T. Flanagan$^2$, E. Schierle$^{1}$, A. Richardella$^2$,  E. D. L. Rienks$^{1,3,4}$,  P. S. Mandal$^{1}$, A. Varykhalov$^{1}$,  J. S\'anchez-Barriga$^{1}$,  F. Radu$^{1}$,   N. Samarth$^2$,  O. Rader$^{1}$}  
\affiliation{	$^1$Helmholtz-Zentrum Berlin f\"ur Materialien und Energie, 
			Elektronenspeicherring BESSY II, Albert-Einstein-Stra\ss e 15, 12489 Berlin, Germany}
\affiliation{$^2$  Department of Physics, Pennsylvania State University, University Park, PA 16802, USA}
 \affiliation{$^3$  Institut f\"ur Festk\"orperphysik, Technische Universit\"at Dresden, 
			01062 Dresden, Germany  }
\affiliation{	$^4$   Leibniz-Institut f\"ur Festk\"orper- und Werkstoffforschung Dresden, 
			Helmholtzstra\ss e 20, 01069 Dresden, Germany}

\date{\today}
 
\def\Tc{$T_{\rm C}$}
\def\BiTe{Bi$_2$Te$_3$}
\def\BiSe{Bi$_2$Se$_3$}
\def\SbTe{Sb$_2$Te$_3$}
\def\BiSbTe{(Bi,Sb)$_2$Te$_3$}
\def\BiSbTe{(Bi, Sb)$_2$Te$_3$}
\def\SbVTe{(Sb$_{1-x}$V$_x$)$_2$Te$_3$}
\def\SbCrTe{(Sb$_{1-x}$Cr$_x$)$_2$Te$_3$}
\def\BiMnTe{(Bi$_{1-x}$Mn$_x$)$_2$Te$_3$}
\def\BiMnSe{(Bi$_{1-x}$Mn$_x$)$_2$Se$_3$}
\def\BiBiMnTe4{Bi$_2$MnTe$_4$}
\def\BiBiMnSe4{Bi$_2$MnSe$_4$}
\def\HgMnTe{Hg$_{1-x}$Mn$_x$Te}
\def\Gbar{\overline{$\Gamma$}}
\def\MnBi{Mn$_{\rm Bi}$} 
\def\muB{$\mu_{\rm B}$}
 
\def\invA{\AA$^{-1}$}
\def\Ef{$E_{\rm F}$}

\begin{abstract}
{\bf V-doped \BiSbTe\ has a ten times higher magnetic coercivity than its Cr-doped counterpart and therefore is
believed to be a superior system for    the quantum anomalous Hall effect (QAHE). 
The QAHE requires the opening of a magnetic band gap at the Dirac point. We do not find this gap by 
angle-resolved photoelectron spectroscopy down to 1 K.
By x-ray magnetic circular dichroism (XMCD) we directly probe the magnetism at the V site and in zerofield. Hysteresis curves of the 
XMCD signal show a strong dependence of the coercivity on the ramping velocity of the magnetic field. 
The XMCD signal decays on a time scale of minutes which we conclude contributes to the absence of a detectable magnetic gap at the Dirac point. 
 }  
 
\end{abstract}

\maketitle

Magnetically doped topological insulators are expected to give rise to a range of new phenomena due to the opening of a magnetic gap at the Dirac point of the topologically protected surface state. 
The quantum anomalous Hall effect, demonstrated in 2013 \cite{Chang13}, is characterized by a quantized Hall resistance  $\rho_{xy} = h/(Ne^2)$  with $N$ an integer number of edge channels. These permit lossless   charge transport in Cr- and V-doped \BiSbTe\ 
\cite{Onoda03,YuScience10,Chang13,CheckelskyNP14,KouPRL14,BestwickPRL15,Kandala2015,ChangCZNM15,GrauerPRB15}. 
%
%
 V-doped \BiSbTe\ was early on found to reach high Curie temperatures of 17 K 
\cite{Zhoua06} and even much higher temperatures were reported for metallic  V concentrations such as 17\%\ V in  \SbTe\
(up to 177 K    \cite{ChienThesis}). 
One advantage of V is that despite its relatively small magnetic moment, its \Tc\ is high due to the high density of states at the
Fermi level meaning a high exchange integral  \cite{VergnioryPRB14}.
A high V density of states was found near the Fermi energy (0.1 to 0.2 eV binding energy \cite{KriegerPRB17}),   using resonant photoemission at the V $L_3$ edge \cite{Peixoto16, KriegerPRB17}. 
By resonant angle-resolved photoemission (ARPES), the V contribution was identified as nondispersing impurity band and found consistent with calculations for V occupying  Bi-substitutional sites \cite{KriegerPRB17}. 

Another advantage is the coercivity $H_{\rm c}$. It is $\sim0.1$ T for Cr-doped \BiSbTe\ 
\cite{Chang13,CheckelskyNP14,KouPRL14} while for the V-doped system  $H_{\rm c}\sim1.3$ T ($T=2$ K) \cite{ChangCZNM15}.
The high coercivity     is believed to lead to a very homogeneous
ferromagnetism with much smaller number of domains that leads to perfect edge state formation \cite{ChangCZNM15}. 
This property facilitated the 
achievement of perfect quantization of $\rho_{xy}$ with a precision of $\pm6\times10^{-4}$ $e^2/h$ 
 and   $\rho_{xx}$ as low as $3.35\pm1.76$ $\Omega$ \cite{ChangCZNM15}. 

%

  
On the other hand, the \Tc\ appears not to be the limiting factor since  the QAHE was observed only at orders of magnitude lower temperatures than the principle limit set by \Tc\ \cite{ChangCZNM15,GrauerPRB15},
even when considering the highest temperatures around 1 K achieved with the help of a depth-dependent concentration profile of Cr impurities \cite{MogiAPL15}.
A decisive parameter is the magnetic band gap at the Dirac point which is caused by the exchange interaction.  
For V-doped \SbTe,  Landau level spectroscopy has been performed by scanning tunneling spectroscopy at 1.3 K \cite{SessiNC16}.   
The method requires the application of high magnetic fields, hence a comparison of samples with and without V doping was used to derive  a mobility gap which amounts to 32 meV. 
This gap does, however, not appear  in the local density of states, possibly due to overlap with impurity states \cite{SessiNC16}. 
  For Cr-doped \BiSbTe, an average  gap of  56  meV was found by tunneling spectroscopy \cite{LeePNAS15}. Its origin could not   be related to magnetism because only data at a single temperature of 4.5 K 
	are available. 
	In ARPES, a large gap at the Dirac point of $\sim$75 meV was found for Cr-doped \BiSe\ even at room temperature, i. e., far above \Tc\ \cite{ChangCZARPESPRL14}.  
 These gaps are therefore not of magnetic origin. 
In a V$_{0.04}$Bi$_{1.96}$Se$_3$ epitaxial film,   ARPES at 100 K showed the opening of a
  gap in the surface state of 180 meV at the Dirac point at $\sim0.3$ eV \cite{ZhangL17}. It was concluded that this gap is not of magnetic 
	origin because a ferromagnetic Curie temperature of 10 K was measured \cite{ZhangL17}.  
The composition of insulating  (Bi, Sb)$_2$(Te, Se)$_3$  can vary strongly between epitaxial films and 
single crystals. This holds also for V-doped  systems  where the composition V$_{0.015}$Bi$_{1.985}$Se$_{0.6}$Te$_{2.4}$ has  the Dirac point at 0.33  eV and V$_{0.03}$Bi$_{1.97}$Se$_{0.6}$Te$_{2.4}$   at 0.38  eV binding energy \cite{Riha20}. 
ARPES of an insulating V-doped \BiSbTe\ epitaxial films showed that the Dirac point is degenerate with the bulk valence band and it was concluded that this limits the achievable temperature of the quantum anomalous Hall effect \cite{Li16}. No magnetic gap could be found in this system, 
(Bi$_{0.29}$Sb$_{0.71}$)$_{1.89}$V$_{0.11}$Te$_3$, down to 7 K \cite{Li16}.

In the present work we undertake a search for the magnetic gap of V-doped (Bi, Sb)$_2$Te$_3$  at the Dirac point down to 1 K. Moreover, we investigate  the magnetization behavior directly and element specifically through magnetic x-ray dichroism in soft x-ray absorption and reflectivity.

Samples were grown by molecular beam epitaxy (MBE) from single elemental sources of at least 5N purity. 
The InP(111) substrate was mounted by indium on the MBE sample holder. 
A  V-doped (Bi, Sb)$_2$Te$_3$  sample of 10 quintuple layer (QL) thickness was grown at 240$^\circ$C with a growth rate of about 0.5 QL per minute. 
 The sample  was capped with 2 nm Te for protection against ambient influence and to allow in situ   removal 
of the cap for ARPES.  The samples have been cut into pieces with a diamond cutter for the various experimental methods.
For transport, magnetization, and soft-x-ray experiments, the cap was kept in place. For the ARPES
experiment, the cap was removed in situ by noble gas ion sputtering and annealing. 
Photoemission experiments were performed with the ARPES-1$^3$ end
station at the UE112-PGM2b undulator beam line of the BESSY II synchrotron
radiation source. In this instrument, the sample can be cooled to 1 K. 
The experimental geometry is the following: With
the central axis of the analyser lens and the polar rotation axis of the
sample defined as the $x$ and $z$ axes of a spherical coordinate system, the
photon beam is incident on the sample under an azimuthal angle   of 45$^\circ$ and a
polar angle of 84$^\circ$. The light polarization is horizontal (along the $x$ axis).
The entrance slit of the hemispherical analyser is placed parallel to the
$z$ axis. The measurements at $h\nu = 105$ eV were performed with an energy
resolution of 10 meV.  
X-ray magnetic circular dichroism (XMCD) was performed at the   XUV diffractometer of the UE46-PGM1 undulator beam line of BESSY II.  
Spectra were taken around the V-L$_{2,3}$ excitation energy for incident photons with opposite circular polarization
and recording the total electron yield through the sample current. 
An ohmic magnet was used which allows also measurements in zero field. 
 
Concerning transport experiments, we see that a square-type hysteresis appears in the transversal magnetoresistance at 4.5 K, Fig. 1(b). 
The longitudinal magnetoresistance, Fig. 1(a), shows peaks at the coercivity typcial for a QAHE sample above
the temperature required for perfect quantization. 
 From  hysteresis measurements in magnetotransport, enlarged in Fig. 1(c), we derive a Curie temperature of 25 K 
in agreement with additional superconducting quantum interference device measurements. 
 
 ARPES data at 1, 20, and 40 K cover the range well below and well above \Tc\ as determined from Fig. 1.
 The angle (respectively ${\bf k}_\parallel$) dependence as well as the normal emission (${\bf k}_\parallel=0$ \invA) spectrum display an intensity maximum at $\sim0.15$ eV binding energy. 
This is due to matrix element effects and can approximately be explained as final-state effect when sampling of the bulk band structure with   varying photon energy. 
The magnetic band gap at the Fermi level requires magnetization perpendicular to the surface plane
\cite{QiPRB08} and scales with the magnetization \cite{Rosenberg12}.
Due to the insulating character, the Dirac point is expected to be at the Fermi level.  
Nevertheless, no gap appears in the data of Fig. 2 with an upper limit of $\sim5$ meV.
In order to independently estimate the position of the Dirac point, we analyzed the width of the momentum distribution curves
in Fig. 2(b) and (d) and arrived at $E_{\rm D} \approx 50$ meV. This indicates a small extent of band bending and  is interestingly  in good agreement with the result by Li et al. for insulating samples who
	located the Dirac point at $54$ meV binding energy \cite{Li16}. 
	 
X-ray magnetic circular dichroism is the method of choice to derive element specific magnetization.
It does not require to take into account transport effects  as in magneto transport or sample and substrate diamagnetism  as in SQUID. 
The XMCD spectrum in Fig. 3(a) displays the L$_2$ and L$_3$ edges which are not fully separated for V. 
The spectrum resembles the one obtained from bulk crystals of  V-doped \SbTe\ \cite{SessiNC16}. 
In  Fig. 3(b--f)   the XMCD signal  at the V $L$-edge is measured as a function of magnetic field. 
 It shows $M(H)$ curves taken at different velocities of 0.005, 0.01, 0.02, 0.038 kG/s. At the slowest speed of 0.005 kG/s, the measured coercivity $H_{\rm c}$ is 0.17 kG.  Surprisingly,  $H_{\rm c}$ increases with increasing speed up to 0.5 kG  at 0.038 kG/s. 
Apparently, the magnetization decays with time. 
Figure 3(g) shows that this occurs  on a time scale of minutes. 

These observations can be related to the magnetism of 
 Cr-doped \BiSbTe\    which is complex and characterized by superparamagnetism 
\cite{Lachman16}.
A strong  influence of the gate voltage on the magnetic relaxation was observed and interpreted as due to the role of charge carriers for the magnetic interaction \cite{Lachman16}.  In that case, the  instability of the magnetization  could be connected to the insulating state. 
On the other hand, the ferromagnetism of   V-doped \BiSbTe\  is much more robust with $10\times$ higher coercivity than for Cr-doping. 
Indeed, the  carriers appear to have no influence on the coercivity as seen from the dependence of the
$\rho_{xy}(H)$ hysteresis on the concentration of Bi from $x=0$ to $x=0.46$ \cite{ChangCZNM15}. 
However, also in  V-doped \BiSbTe, electronic and magnetic effects have been found to be connected 
\cite{GrauerPRB15}. 
  It is unclear how the observed self-magnetization effect \cite{ChangCZNM15} can play a role here:
	It was observed that in contrast to Cr-doped \BiSbTe, the V-doped system displays a QAHE in zerofield.
	In the hysteresis curves $\rho_{xy}(H)$ the virgin curve is not different from the hysteresis and starts at
	$H=0$ with the saturation value. 
	It is not established whether this occurs only in transport or also in conventional magnetometry.
	
The observed decay of the magnetization can directly affect the ARPES data which are measured on a much longer timescale. 
It may be that the magnetization decay prohibits the observation of the magnetic gap at the Dirac point. 
On the other hand,  a magnetic gap of up to 90 meV  can be measured by ARPES  in Mn-doped \BiTe\ 
   below \Tc\ \cite{Rienks18}.  
There are, however, important differences between the samples. The first one is the insulating character which does not apply to the strongly n-doped Mn-doped \BiTe.
Another one is that the Mn-doped system avoids disorder by forming a more uniform septuple/quintuple layer 
heterostructure \cite{Rienks18}. 
In addition to the decreased disorder, this structure also enhances the exchange interaction so that the measured
gap is about five times larger than predicted for a comparable dilute system  with the same Mn concentration.
  V-doped \BiSbTe\ is such a dilute substitutional system and the gap may just be too small to observe in ARPES.
	The finding of the bulk valence band at the Fermi energy \cite{Li16} may explain the limitation of the QAHE to 
low temperature. It cannot explain the absence of the gap at the Dirac point since it is clearly separated in 
ARPES data by the electron momentum parallel to the surface. 
Finally, the magnetism of Mn is much more localized than the one of V and this difference may also be relevant in the dilute system. For such reason, in elemental Fe the exchange splitting as measured by ARPES persists above \Tc\ whereas it vanishes in Ni. 
In those elemental ferromagnets  the exchange splitting is  maintained below \Tc\ in magnetic domains even if the macroscopic magnetization is zero. 
From recent work on Cr-doped   \BiSbTe\  \cite{Lachman16} we learn, however, that the domain structure is not at all of the type well known from  elemental Fe and Ni and this may the reason why the decay of magnetization below \Tc\ 
leads to the absence of a measurable magnetic gap at the Dirac point. 

In summary, we have investigated V-doped   \BiSbTe\ system which is close to the QAHE by synchrotron radiation methods ARPES and XMCD. 
We find that a small downward band bending occurs but no magnetic gap can be observed at the Dirac point down to T = 1 K, much below the Curie temperature of $\sim25$ K. 
  The XMCD experiments which give access to  the element specific magnetization of the 
V  reveal a surprising decay of the magnetization in zerofield. This could prevent the observation of the magnetic gap at the Dirac point by ARPES. 
It would be very helpful to gain insight into the magnetic domain structure at temperatures of 1 K and below.

\begin{figure}
\centering
\includegraphics[width=0.55\linewidth]{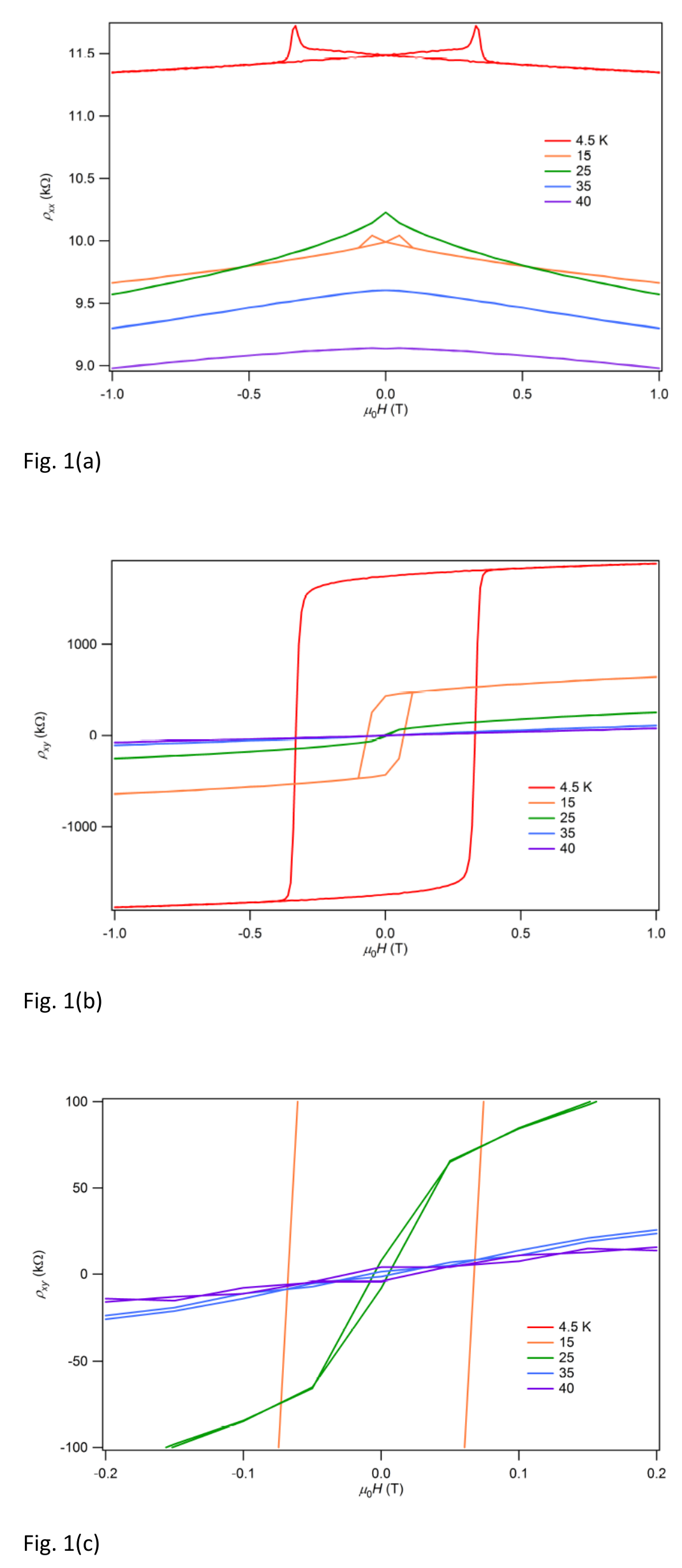}
\caption{(a) Longitudinal and (b,c) transversal resistance vs. magnetic field.  A \Tc\
of 25 K is derived.} 
\end{figure}
 
\begin{figure}
\centering
\includegraphics[width=0.7\linewidth]{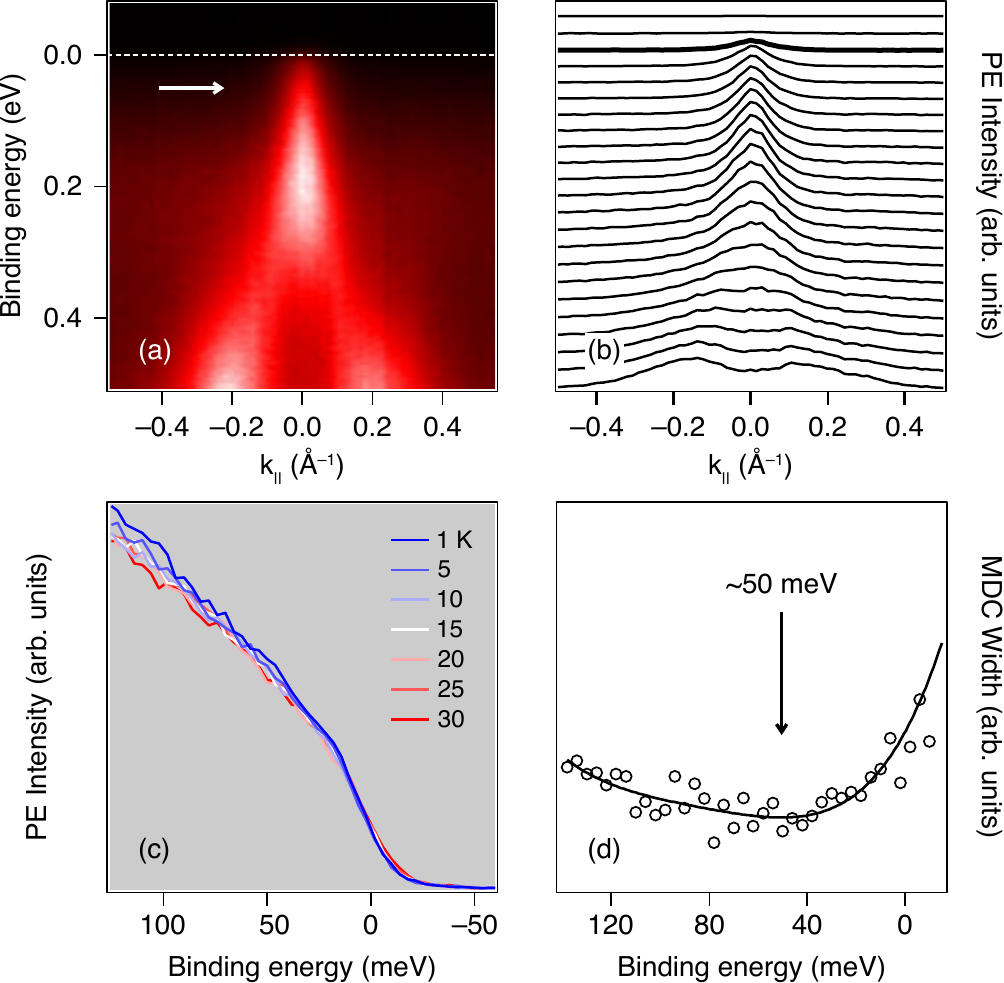}
\caption{ARPES spectra at (a,b) 5 K and (c) the temperature dependence in normal emission ($k_\parallel=0$ \AA$^{-1}$)
covering the range above and below the ferromagnetic phase transition.   No apparent gap occurs down to 1 K. (d) The position of the Dirac point is determined from the width of the momentum distribution curves (MDC) in (b) as $-50$ meV. This position is marked by arrows in panels (d) and (a).}
\end{figure}
 
\begin{figure}
\centering
\includegraphics[width=0.55\linewidth]{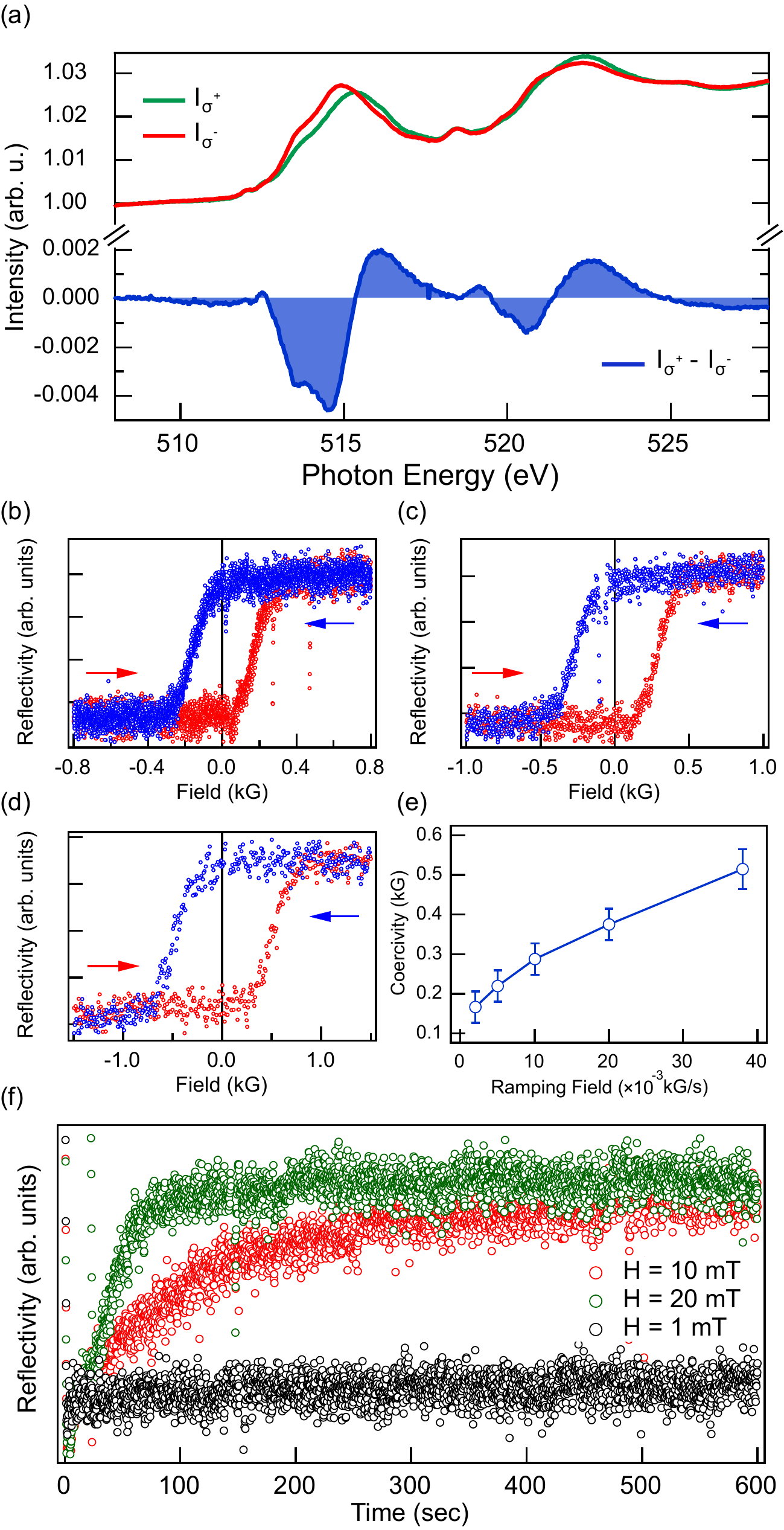}
\caption{(a) X-ray absorption in total electron yield and x-ray
magnetic circular dichroism (XMCD) at 4 K. Time dependence of XMCD
hysteresis loops at 4 K (b--f). Plot of $H_{\rm c}$ vs. $v$ (g).
Plot of time decay of XMCD signal (h).} \end{figure}

\noindent {\bf Acknowledgements}

\noindent {O. R. thanks Andreas Ney for helpful discussions. This work was supported by   SPP1666 of Deutsche Forschungsgemeinschaft. }

\noindent {\bf Data availability}

\noindent The data that support the findings of this study are available from the corresponding author upon reasonable request.


\end{document}